# Deterministic Annealing Optimization for Witsenhausen's and Related Decentralized Stochastic Control Problems

Mustafa Said Mehmetoglu, *Student Member, IEEE,* Emrah Akyol, *Member, IEEE,*
and Kenneth Rose, *Fellow, IEEE*

*Abstract*—This note studies the global optimization of controller mappings in discrete-time stochastic control problems including Witsenhausen's celebrated 1968 counter-example. We propose a generally applicable non-convex numerical optimization method based on the concept of deterministic annealing – which is derived from information theoretic principles and was successfully employed in several problems including vector quantization, classification and regression. We present comparative numerical results for two test problems that show strict superiority of the proposed method over prior approaches in literature.

## I. INTRODUCTION

Decentralized control systems have multiple controllers designed to collaboratively achieve a global objective while taking actions based on local observations. One of the most studied structures, termed "linear quadratic Gaussian" (LQG), involves linear dynamics, quadratic cost functions and Gaussian variables. Since in the case of centralized LQG problems, the optimal mappings are linear, it was naturally conjectured that linear control mappings remain optimal in decentralized settings. However, Witsenhausen proposed an example of a decentralized LQG control problem, commonly referred to as Witsenhausen's counter-example (WCE), for which he provided a simple non-linear control strategy that outperforms all linear strategies [1]. The problem has been viewed as a benchmark in stochastic networked control, see, e.g., [2] for a detailed treatment.

Decentralized control systems such as WCE arise in many practical applications, and numerous variations on WCE have been studied in the literature (see, e.g., [3]–[8]). In general, linear control strategies are not optimal for LQG systems, except when the system admits some specific information structure (see, e.g., [9], [10]). It is well-understood that if the information structure in a decentralized control problem is nonclassical, as in the case of WCE, non-linear strategies may widely outperform optimal linear strategies. Finding the optimal mappings for such problems is usually a difficult task unless they admit an explicit (and often as simple as linear) solution [3].

Recent research efforts have focused on developing efficient numerical methods for decentralized control problems [11]–[14], specifically for WCE [15]–[18]. Some of the existing methods rely on simplifying properties of WCE, such as monotonicity, and are therefore not easily generalizable [11], [12], [19]. Moreover, methods that require analytical derivation for each particular setting are not fully automated [14], [19].

In this work, building on our prior work [20], we propose an optimization method, based on the concept of deterministic annealing (DA), for a class of decentralized stochastic control problems. DA has been successfully used in various problems in control theory including dynamic coverage control problems [21]–[23] and cluster analysis in control systems [24], in addition to other problems involving non-convex optimization such as vector quantization [25], regression [26], zero-delay source-channel coding [27], and more (see review in [28]). There are many important advantages of the proposed method compared to prior work, including ability to avoid poor local minima and independence from initialization.

We demonstrate the DA-based method on two specific problems. We first analyze the numerically "over-mined" WCE problem. We then study a more involved variation, introduced in [7], which includes an additional noisy channel over which the two controllers communicate. The second controller, therefore, has access to some side information which is controlled by the first controller. We refer to this setting as the "side channel problem" motivated by the class of "decoder side information" problems in communications and information theory [29]. It has been demonstrated in [7] that non-linear strategies may outperform the best linear strategies, however, the question of how to approach the optimal solution remains open.

Having a powerful optimization method at hand, we analyze the structure of experimentally obtained mappings. For instance, Wu and Verdú have shown [30] that the optimal solution of WCE must have real analytic left inverse, thus, a piecewise linear function cannot be optimal. Our numerical results demonstrate that the "steps" in obtained mappings show small deviations from linear, experimentally confirming this theoretical finding.

Mustafa S. Mehmetoglu and Kenneth Rose are with the Department of Electrical and Computer Engineering, University of California, Santa Barbara, CA, 93106, USA (e-mail: mehmetoglu@ece.ucsb.edu, rose@ece.ucsb.edu)

Emrah Akyol is with the Coordinated Science Laboratory, University of Illinois at Urbana-Champaign, Urbana, IL, 61801, USA (email: akyol@illinois.edu)

This work is supported by the NSF under grants CCF-1118075 and CCF-1016861. The material in this paper was presented in part at the IEEE International Symposium on Information Theory (ISIT), Honolulu, HI, USA, June 2014

## II. PROBLEM DEFINITION

### A. Notation

Let $\mathbb{R}$, $\mathbb{E}(\cdot)$ and $\mathbb{P}(\cdot)$ denote the set of real numbers, the expectation and probability operators, respectively. We represent random variables and their realizations with uppercase and lowercase letters (e.g., $X$ and $x$), respectively. Let $X_i^j$ denote the set $X_i, \ldots, X_j$. The probability density function of the random variable $X$ is $f_X(x)$. The Gaussian density with mean $\mu$ and standard deviation $\sigma$ is denoted as $\mathcal{N}(\mu, \sigma^2)$. We use natural logarithms which, in general, may be complex, and the integrals are, in general, Lebesgue integrals.

### B. General Problem Definition

Formally, we consider a discrete-time stochastic control problem with nonclassical information pattern involving $n$ controllers, and assume that the order of control actions is fixed in advance, i.e., the system is sequential [31]. Following the problem definition in [31], let $(\Omega, \mathcal{B}, \mathcal{P})$ be a probability space, where $\Omega$ denotes the random quantities involved in the system such as initial input, and $(U_i, \Sigma_i)$, for $i = 1, \ldots, n$, are measurable spaces with $U_i$ denoting the set of control actions. Controller mappings (functions) are denoted by $g_i : \Omega \times U_1 \times U_2 \ldots U_{i-1} \to U_i$, for $i = 1, \ldots, n$. For convenience, we denote the input set of $g_i$ by $X_i = \Omega \times U_1 \times U_2 \ldots U_{i-1}$. The system is then defined by the following set of equations

$$u_i = g_i(x_i), \ i = 1, \ldots, n. \quad (1)$$

Let $f$ be a real-valued and bounded measurable function of $\omega, u_1^n$ on $(\Omega, \mathcal{B})$, i.e., $f$ is a random variable. The problem objective is to find the set of functions $g_1^n$ that minimize the value of the cost function $J$:

$$J = \mathbb{E}\{f(\omega, u_1^n)\}. \quad (2)$$

## III. PROPOSED METHOD

A quick overview of the proposed method at the high level is as follows: We introduce controlled randomization into the optimization process by randomizing the controller mappings, and impose a constraint on the level of randomness (measured by the Shannon entropy) while minimizing the expected cost of the system. The resultant Lagrangian functional can be viewed as the "free energy" of a corresponding physical system, wherein the Lagrangian parameter is the "temperature". The optimization is equivalent to an annealing process that starts by minimizing the cost (free energy) at a high temperature, which effectively maximizes the entropy. The minimum cost is then tracked at successively lower temperatures as the system typically undergoes a sequence of phase transitions through which the complexity of the solution (controller mappings) grows. As the temperature approaches zero, hard (nonrandom) mappings are obtained. DA avoids poor local minima through randomization of mappings, slowly tracks the minimum cost, and can achieve the global minimum under certain conditions on the type (and continuity) of phase transitions. However, there is no general guarantee of convergence to the globally optimal solution.

### A. Derivation

Let us denote the space of controller input $x_i$ by $\mathbb{R}_{x_i}$, for $i = 1, \ldots, n$. Assume there exists a partition of $\mathbb{R}_{x_i}$ into $\mathcal{M}_i > 0$ disjoint regions denoted by $\mathbb{R}_{i,m_i}$ ($m_i = 1, \ldots, \mathcal{M}_i$):

$$\bigcup_{m_i=1}^{\mathcal{M}_i} \mathbb{R}_{i,m_i} = \mathbb{R}_{x_i}. \quad (3)$$

Note that each value of $x_i$ belongs to exactly one of the partition regions, referred to as a deterministic (non-random) partition.

We begin our formulation by imposing a piecewise structure on the controller mappings. Consider the structured mapping $g_i$, for $i = 1, \ldots, n$, written as

$$g_i(x_i) = g_{i,m_i}(x_i) \quad \text{for } x_i \in \mathbb{R}_{i,m_i}. \quad (4)$$

Each $g_{i,m_i}(x_i)$ is a parametric function referred to as "local model". Effectively, each of the mappings $g_i$ is defined with a structure determined by two components: a space partition where regions are denoted by $\mathbb{R}_{i,m_i}$ and a parametric local model per partition cell, i.e., $g_{i,m_i}(x_i)$ for $\mathbb{R}_{i,m_i}$. The number of local models (partition regions) for mapping $g_i$ is $\mathcal{M}_i$. The local models can take any prescribed form such as linear, quadratic or Gaussian and we let $\Lambda(g_{i,m_i})$ denote the parameter set for local model $g_{i,m_i}$.

The crucial idea in DA is to introduce controlled randomization into the problem formulation. We replace the deterministic partition of space by a random partition, i.e., we associate every input point ($x_i$) with partition regions *in probability*. To this end, we introduce random variables $M_i$, for $i = 1, \ldots, n$, whose realization is the partition index $m_i$. We define the *association probabilities* as conditional distribution on the partition index given the input:

$$p_i(m_i|x_i) = \mathbb{P}\{x_i \in \mathbb{R}_{i,m_i}\} = \mathbb{P}\{g_i(x_i) = g_{i,m_i}(x_i)\}, \quad (5)$$

for $i = 1, \ldots, n$. Consequently, the mappings are now random, in the sense that the output of controller $g_i$ for an input $x_i$ is given in probability as

$$g_i(x_i) = g_{i,m_i}(x_i) \quad \text{with probability } p_i(m_i|x_i). \quad (6)$$

By construction, we have that given $X_i$, $M_i$ is independent of the random variables $M_1^{i-1}$.

The expectation in (2) is now taken over $X_i^n$ and $M_i^n$. Let us rewrite it for a fixed value of $i$ as follows:

$$J = \int_{x_i} \sum_{m_i=1}^{\mathcal{M}_i} \mathbb{E}\{f(\omega, u_1^n)|m_i, x_i\} p_i(m_i|x_i) f_{x_i}(x_i) \mathrm{d}x_i \quad (7)$$

where $\mathbb{E}\{f(\omega, u_1^n)|m_i, x_i\}$ can be viewed as the cost of associating $x_i$ with local model $g_{i,m_i}$. Assuming fixed local model parameters, optimizing (7) with respect to $p_i(m_i|x_i)$ would clearly produce deterministic mappings, since the minimum is achieved by setting $p_i(m_i|x_i) = 1$ for the pair $\{m_i, x_i\}$ for which $\mathbb{E}\{f(\omega, u_1^n)|m_i, x_i\}$ is minimum. Therefore, the ultimate objective of obtaining optimum deterministic controllers is preserved as the random encoders share the same global minimum as deterministic ones. However, direct optimization of the cost with respect to $p_i(m_i|x_i)$ results in poor local minima. Instead, we minimize (2) at prescribed levels of *randomness*, which we measure by the Shannon entropy. The joint entropy of the system can be written

$$H(X_1^n, M_1^n) = H(X_1) + \sum_{i=2}^{n} H(X_i|M_1^{i-1}, X_1^{i-1})$$
$$+ H(M_1|X_1) + \sum_{i=2}^{n} H(M_i|X_i, M_1^{i-1}, X_1^{i-1}). \quad (8)$$

It is easy to see from conditional independence arguments that the conditional entropies in the second term in the righthand side of (8) can be simplified to $H(X_i|X_1^{i-1})$, and those in the last term to $H(M_i|X_i)$. Thus the first two terms of (8) are fixed and determined by the problem statement (the joint distribution of $X_1^n$). We therefore discard the first two fixed terms of (8), rearrange the remaining terms, to obtain a conveniently compact measure of randomness defined as

$$H \triangleq \sum_{i=1}^{n} H(M_i|X_i). \quad (9)$$

The conditional entropy $H(M_i|X_i)$ is given by

$$H(M_i|X_i) = -\int_{x_i} \sum_{m_i=1}^{\mathcal{M}_i} p(m_i|x_i) \log p(m_i|x_i) f_{X_i}(x_i) \mathrm{d}x_i. \quad (10)$$

Accordingly, we construct the Lagrangian

$$F = J - TH \quad (11)$$

as the objective function to be minimized, where $J$ is given in (2), $H$ is given in (9) and $T$ is the Lagrange multiplier associated with the entropy constraint. The notational choices are to emphasize the analogy to statistical physics, where $F$ can be viewed as the Helmholtz free energy, $J$ as the thermodynamic energy, $H$ as the entropy, and $T$ as the temperature, of a corresponding physical system (see [28] for a detailed treatment of the statistical physics analogy).

### B. Optimization Method

The practical method consists of gradually reducing the temperature $T$ while tracking the minimum of the free energy $F$. The central iteration at each temperature consists of optimizing the local model parameters and association probabilities. Initially, for $T \to \infty$, the minimum $F$ is obtained by maximizing $H$, which is achieved by uniform association probabilities as can be seen from (9). Consequently, for each controller, the local models are all optimized for the same distribution over the input space, and are therefore all identical, i.e., there is effectively a single "distinct" local model. As the temperature is decreased, the system will undergo "phase transitions" where the current solution no longer represents a minimum (the old minimum typically becomes a saddle point) and there exists a better solution with increased number of distinct local models. Since it is computationally more efficient to keep only distinct models, we initialize with a single model ($\mathcal{M}_i = 1$) and trigger phase transitions by duplicating existing models and slightly perturb them at each temperature. The minimization of $F$ will either produce the existing solution (the duplicated and perturbed local models seek to merge at their initial values), or a better solution with increased number of distinct models, depending on whether a "critical temperature" has been reached. Accordingly, in we combine identical models. At the limit $T \to 0$, we perform zero temperature iteration (equivalent to gradient descent) by fully assigning source points to the model that makes the smallest contribution to $J$, thus obtaining the desired deterministic mappings. (This is referred to as "quenching" in the physical analogy.)

*Remark 1:* Our method is derived without recourse to discretization. Although practical simulations involve sampling of the continuous space during numerical computations of integrals, this is in contrast to methods that are entirely formulated in discrete settings.

*Remark 2:* Critical temperatures can be derived analytically if, for the problem considered, phase transitions are of "continuous" nature, in the sense that tracked minimum becomes a saddle point at the exact critical temperature. The condition for saddle point can be obtained using variational calculus, see [28] for phase transition analysis in DA. Our experiments indicate that, at least for the test cases considered in this paper, phase transitions are not continuous. While pre-calculating the critical temperature may enable a numerical speed up of the annealing process, it is not necessary to implementing the practical algorithm. Hence, the derivation and characteristics of phase transitions are kept outside the scope of this paper.

The optimal $p_i(m_i|x_i)$ that minimize (11) can be derived in closed form. Plugging (7) and (10) in (11), straightforward derivation gives the optimal $p_i(m_i|x_i)$ as

$$p_i(m_i|x_i) = \frac{e^{-\mathbb{E}\{f(\omega, u_1^n)|m_i, x_i\}/T}}{\sum_{m_i} e^{-\mathbb{E}\{f(\omega, u_1^n)|m_i, x_i\}/T}} \quad (12)$$

Optimization of parameters in $\Lambda(g_{m_i})$ can be done using any standard method. Typically, a variant of gradient descent is used when closed form expressions cannot be obtained.

## IV. APPLICATIONS OF THE PROPOSED DA METHOD

### A. Witsenhausen's Counterexample

*1) Problem Description:* Let $X_0$ and $W$ be Gaussian random variables with distributions $\mathcal{N}(0, \sigma_{X_0}^2)$ and $\mathcal{N}(0, 1)$, respectively. WCE is a 2-stage control problem with controllers $g_1 : \mathbb{R} \to \mathbb{R}$ and $g_2 : \mathbb{R} \to \mathbb{R}$, defined by the following equations:

$$U_1 = g_1(X_0), \quad U_2 = g_2(X_1 + W),$$
$$X_1 = X_0 + U_1, \quad X_2 = X_1 - U_2. \quad (13)$$

The schematic representation is given in Figure 1a. The objective is to minimize the cost

$$J = \mathbb{E}\{k^2 U_1^2 + X_2^2\}. \quad (14)$$

For convenience, we define $f_1(X_0) = g_1(X_0) + X_0$.

Some properties of $f_1(\cdot)$ are known, including the property of symmetry about the origin (thus, positive half is enough to describe a given solution) [1]. Witsenhausen has provided the following solution that outperforms the optimal linear solution for a given set of problem parameters ($k = 0.2$, $\sigma_{X_0} = 5$):

$$f_1(x_0) = \sigma_{X_0} \operatorname{sgn}(x_0) \tag{15}$$

where $\operatorname{sgn}(\cdot)$ is the signum function. Since there is a single "step" in the positive half of real line, this solution is referred to as a "1-step" solution. Improved solutions that appeared in literature utilize 2.5, 3, 3.5 and 4-step functions (an $x.5$ step function has a step that straddles the origin). Moreover, the latter solutions made improvements by using slightly sloped steps rather than constant ones.

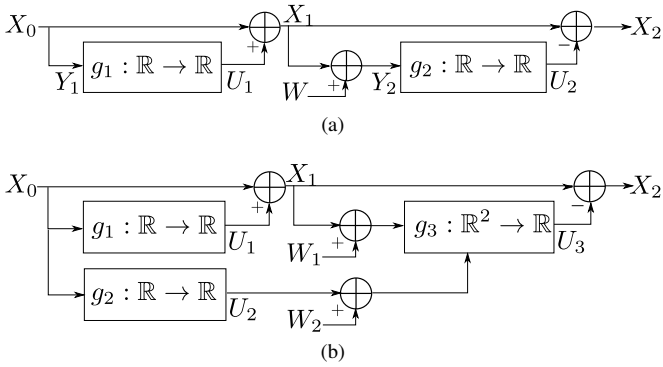

Fig. 1. Settings used for testing the proposed method (a) Witsenhausen's counter-example (b) Side channel problem.

TABLE I
RESULTS FOR WCE

| Solution | Cost |
|---|---|
| Optimal linear Solution | 0.96 |
| 1-step, Witsenhausen [1] | 0.404253 |
| 2-step, [11] | 0.190 |
| Sloped 2.5 - step, [15] | 0.1701 |
| Sloped 3.5 - step, [19] | 0.1673132 |
| Sloped 3.5 - step, [17] | 0.1670790 |
| Sloped 4 - step, [16] | 0.16692462 |
| Sloped 5 - step, our result | 0.16692291 |

Although in standard application of DA-based method we randomize all controllers, for computational efficiency, we restrict the randomization to only $g_1$ as

$$g_1(x_0) = g_{1,m_1}(x_0) \quad \text{with probability } p_1(m_1|x_0) \tag{16}$$

and numerically compute (update) $g_2$ by using the fact that optimal $g_2$ given $g_1$ is

$$g_2(Y_2) = \mathbb{E}\{X_1|Y_2\} \tag{17}$$

where $Y_2 = X_1 + W$.

For this particular problem, we use linear local models given by

$$g_{1,m_1}(x_0) = a_{1,m_1} x_0 + b_{1,m_1}. \tag{18}$$

while noting that optimal $g_1$ must have analytic left inverse and hence cannot be piecewise linear [30]. Nevertheless, the minimal cost can be approached arbitrarily closely by piecewise linear functions [30]. Thus, for numerical algorithms, linear models are sufficient.

*2) Results for WCE:* Preliminary results on the application to WCE appeared in [20]. We first provide results for the standard benchmark case where $k = 0.2$, $\sigma_{X_0} = 5$ that was used in many papers in literature. The annealing process is illustrated in Figure 2, where evolution of the mapping can be seen (only the positive half is shown thanks to the symmetry property). The obtained mapping is referred to as a "sloped 5-step" solution. At high temperature, there is only one local model, thus, the function is 1-step. As the temperature is lowered, the solution undergoes phase transitions, revealing more steps for the mapping function. In this work we calculated the solution with 5 steps. Although more steps possibly exist, improvement to cost is numerically insignificant with additional steps. Some earlier results from the literature are given in Table I, where it can be seen that our method produced the minimum cost achieved to date.

Another benchmark case, $k = 0.63$, was suggested in [5] as potentially being more relevant for confirming the high gains of optimal non-linear mappings. Our resulting mapping for $k = 0.63$, $\sigma_{X_0} = 5$ is given in Figure 3a, which is a 6-step solution. Our numerical results suggest that the gain over linear solution is smaller compared to the standard benchmark case above: $J = 0.844$ for the solution in Figure 3a whereas cost associated with the optimal linear mapping is $J = 0.961$.

These numerical results illustrate an important theoretical result as well. In [30] authors proved that the optimal $f_1$ must have analytic left inverse and therefore cannot be piecewise linear, which was believed to be the case due to numerical results (see, e.g., [19]). Our numerical results indicate that steps are in fact non-linear, as shown in Figure 3b. The steps become non-linear during the final stages of the algorithm as multiple local models appear to form a single step. To the best of our knowledge, this is the first numerical result illustrating non-linearity of the steps.

*B. Side Channel Problem*

*1) Problem Description:* Let $X_0$ be a Gaussian random variable with distribution $\mathcal{N}(0, \sigma_{X_0}^2)$, and $W_1$, $W_2$ be independent Gaussian random variables, both with a distribution $\mathcal{N}(0, 1)$. The system is defined by the following equations:

$$U_1 = g_1(X_0), \ U_2 = g_2(X_0), \ U_3 = g_3(X_1 + W_1, U_2 + W_2),$$
$$X_1 = X_0 + U_1, \quad X_2 = X_1 - U_3. \tag{19}$$

The problem is to optimize the cost function

$$J = \mathbb{E}\{k^2 U_1^2 + X_2^2\} \tag{20}$$

for given $\sigma_{X_0}$ and positive parameter $k$, subject to a power constraint on $U_2$:

$$b_{SNR} = \mathbb{E}\{U_2^2\} \tag{21}$$

where $b_{SNR}$ is the specified power level. We again define $f_1(X_0) = g_1(X_0) + X_0$.

This problem setting is illustrated in Figure 1b and was introduced in [7]. It can be seen as a generalization of WCE with an additional communication channel between the controllers, i.e., a non-linear function of input $X_0$ is communicated by $g_2$ to the controller denoted by $g_3 : \mathbb{R}^2 \to \mathbb{R}$. The non-linear mappings analyzed in [7], which widely outperform the best linear solution in a large range of $b_{SNR}$, are such that both $f_1$ and $g_2$ are staircase functions of $x_0$.

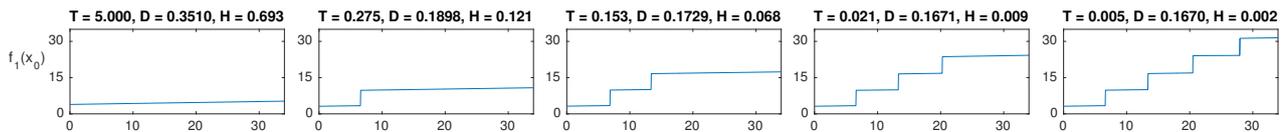

Fig. 2. Evolving graph of $f_1(x_0)$ in WCE during various phases of the annealing process. We note that mapping is actually random during algorithm run. Here, for demonstration, we fully associate every $x_0$ with $g_{1,m_1}$ for which $p_1(m_1|x_0)$ is largest.

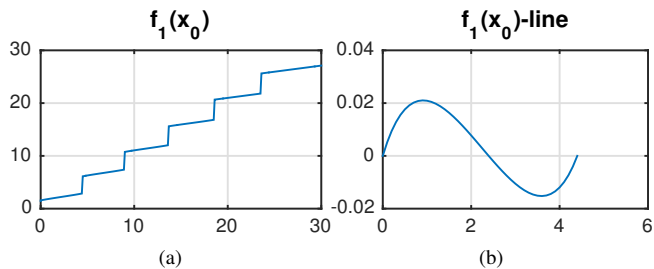

Fig. 3. Numerical result for WCE in the case of $k = 0.63$, $\sigma_{X_0} = 5$. (a) 6-step solution. (b) The deviation of the first step in $f_1(x_0)$ from a straight line between the end points of the step.

TABLE II
COST COMPARISON TABLE FOR SIDE CHANNEL PROBLEM

| $b_{SNR}$ | Linear Cost | $J^M$ ( [7]) | $J^*$ | $(J^M - J^*)/J^M$ |
|---|---|---|---|---|
| 0   | 0.960 | 0.185 | 0.167 | 0.10 |
| 2.6 | 0.696 | 0.149 | 0.079 | 0.47 |
| 4.7 | 0.432 | 0.101 | 0.040 | 0.60 |
| 5.9 | 0.344 | 0.081 | 0.026 | 0.68 |
| 9.0 | 0.203 | 0.052 | 0.012 | 0.77 |

*2) Results for Side Channel Problem:* The original problem is to minimize (20) subject to the constraint in (21). We follow the standard approach in optimization theory and convert this constrained problem to unconstrained Lagrangian formulation:

$$J = \mathbb{E}\{k^2 U_1^2 + X_2^2 + \lambda U_2^2\} \quad (22)$$

where $\lambda$ is chosen to satisfy the power constraint (21) with equality. In the experiments, we used the standard benchmark parameters that were used for the original WCE, that is, $k = 0.2$ and $\sigma_{X_0} = 5$. We have varied $\lambda$ to obtain results at different $b_{SNR}$.

In Table II we compare the cost of our solutions (denoted by $J^*$) to the ones given in [7] (denoted by $J^M$), and the best linear mappings. Significant cost reductions can be observed. The relative improvement over the solution of [7] is listed in the last column.

*Remark 3:* When $b_{SNR} = 0$, the problem degenerates to WCE, thus the cost is 0.1669, the best known to date.

We present several mappings obtained by our method in Figure 4. Some interesting features of these mappings are observed. The mappings $f_1$ are staircase functions with constant steps similar to the ones obtained for the original WCE problem, however, the steps get smaller and increase in number as the side channel SNR increases; that is, $f_1(x_0)$ approaches $x_0$. Note that the control cost term in (20), $\mathbb{E}\{k^2 U_1^2\}$, achieves its minimum when $g_1 = 0$, i.e., $f_1(x_0) = x_0$. This is, however, not optimal due to the estimation error at the second stage. Intuitively, as the second controller has access to better side information (i.e. at higher SNR), the estimation error is decreased and as observed in Figure 4, $f_1(x_0)$ approaches $x_0$. The relative improvement in cost, given in Table II, increases with SNR, which is consistent with the above observation.

The mappings for the side channel, $g_2$, are irregular and the overall shape varies with SNR. This observation, together with the above for $f_1$, suggests that the mappings $f_1$ and $g_2$ are not scale invariant. The discontinuities in $f_1$ and $g_2$ coincide as expected, as the discontinuities in side information signal those in $f_1$ to $g_3$.

**Note**: For numerical results in this paper, MATLAB codes can be found in [32].

## V. ADVANTAGES OF PROPOSED METHOD

There are several improvements of the method proposed here over existing methods in literature.

1) It is derived in the original, continuous domain, without discretization. The continuous space is sampled during numerical computation of integrals only. This is in contrast with many prior methods such as those in [12], [16], [17] that are entirely formulated in a discrete setting.
2) Our method is based on DA, a powerful non-convex optimization framework. DA has been successfully used as a remedy to the problem of poor local minima in non-convex optimization problems [28], and is shown to outperform competing methods such as "noisy channel relaxation" (NCR) [27]. Although NCR performs well for the simple setting of WCE [16], it is susceptible to get trapped in local minima in more involved settings.
3) From its DA foundation, our method directly inherits notably useful properties including reduced sensitivity to initialization. The authors in [15] had to experiment with a large number of initial weight vectors to obtain the result included in Table I.
4) We do not make any assumptions about the controller mappings. Methods presented in [11], [12], [19] benefit from the monotonicity of optimal mapping in WCE. The results in Figure 4 demonstrate that monotonicity may not hold for optimal mappings in the general setting.
5) The method is fully automated and does not require analytical derivations or manual interventions during algorithm run. This is in contrast to the method presented in [19] which requires analytical work during the procedure.
6) Method is applicable to a broad class of stochastic control problems. Many prior methods require non-trivial work in order to generalize, for instance, [17]

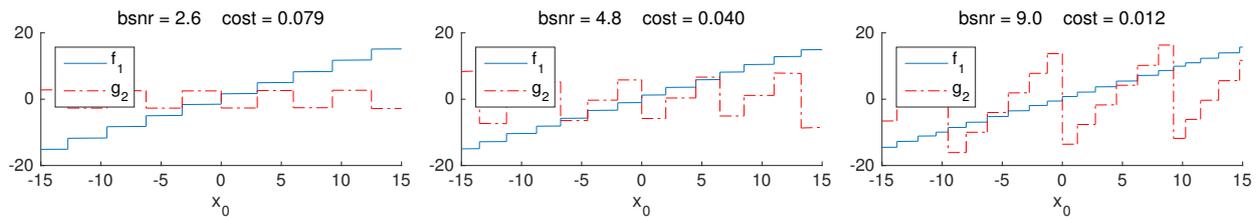

Fig. 4. Some of the mappings we obtained for the side channel variation problem. The first controller is plotted at various SNR levels.

proposes a method that is generalizable, but it requires conversion to a potential game problem.

## VI. CONCLUSIONS

In this paper, we proposed an optimization method for distributed control problems, whose solutions are known to be non-linear, and demonstrated its effectiveness on two problems from the literature. The first problem is the celebrated benchmark problem known as Witsenhausen's counter-example, for which our approach obtained the best known cost value. As a second test case we focused on the side channel setting introduced in [7], where it is motivated as a two stage noise cancellation problem. The mappings obtained are highly nontrivial, offer considerably improved performance, and raise interesting questions about the functional properties of optimal mappings in decentralized control, which are the focus of ongoing research.